# Liquid Metal Printed Ultrathin Oxides for Monolayer WS$_2$ Top-Gate Transistors

*Yiyu Zhang[1], Dasari Venkatakrishnarao[1], Michel Bosman[1,2], Wei Fu[1], Sarthak Das[1], Fabio Bussolotti[1], Rainer Lee[1], Siew Lang Teo[1], Ding Huang[1], Ivan Verzhbitskiy[1], Zhuojun Jiang[1], Zhuoling Jiang[3], Jian Wei Chai[1], Shi Wun Tong[1], Zi-En Ooi[1], Calvin Pei Yu Wong[1], Yee Sin Ang[3], Kuan Eng Johnson Goh[1,4,5,\*], Chit Siong Lau\*,[1]*

**Keywords**: 2D material, dielectric, subthreshold swing, nanoelectronics, 2D semiconductor, interface engineering, gate leakage, equivalent oxide thickness, valley polarization

[1] Institute of Materials Research and Engineering, Agency for Science, Technology and Research (A*STAR), 2 Fusionopolis Way, Innovis, 138634, Singapore

[2] Department of Materials Science and Engineering, National University of Singapore, 9 Engineering Drive 1, 117575, Singapore

[3] Science, Mathematics and Technology, Singapore University of Technology and Design, 8 Somapah Road, 487372, Singapore

[4] Department of Physics, National University of Singapore, 2 Science Drive 3, 117551, Singapore

[5] Division of Physics and Applied Physics, School of Physical and Mathematical Sciences, Nanyang Technological University, 50 Nanyang Avenue 639798, Singapore

\*Email: aaron_lau@imre.a-star.edu.sg; kejgoh@yahoo.com

**Two-dimensional (2D) semiconductors are promising channel materials for continued downscaling of complementary metal-oxide-semiconductor (CMOS) logic circuits. However, their full potential continues to be limited by a lack of scalable high-*k* dielectrics that can achieve atomically smooth interfaces, small equivalent oxide thicknesses (EOT), excellent gate control, and low leakage currents. Here, we report liquid metal printed ultrathin and scalable Ga$_2$O$_3$ dielectric for 2D electronics and electro-optical devices. We directly visualize the atomically smooth Ga$_2$O$_3$/WS$_2$ interfaces enabled by the conformal nature of liquid metal printing. We demonstrate atomic layer deposition compatibility with high-*k* Ga$_2$O$_3$/HfO$_2$ top-gate dielectric stacks on chemical vapour deposition grown monolayer WS$_2$, achieving EOTs of ~1 nm and subthreshold swings down to 84.9 mV/dec. Gate leakage currents are well within requirements for ultra-scaled low-power logic circuits. Our results show that liquid metal printed oxides can bridge a crucial gap in scalable dielectric integration of 2D materials for next-generation nano-electronics.**

Field effect transistors (FETs) are the fundamental building blocks for modern computing processors. The world's need for faster, greener, and more complex capabilities drives continued miniaturization in FETs. Transistors in integrated circuits have doubled about every two years for more than half a century, in a trend known as Moore's law.[1] However, continued downscaling in size of bulk semiconductors faces challenges from issues such as increased scattering from rough channel-insulator interfaces and increased gate leakage currents.[2–4]

One promising solution to allow further FET miniaturization is atomically thin 2D semiconductors with pristine surfaces that lack dangling bonds.[2–8] 2D FETs based on semiconducting transition metal dichalcogenides (TMDCs) like $WS_2$ are prime candidates to extend the scaling roadmap for logic device.[9] Their potential for excellent electrostatic gate control, high carrier mobility, and their unique spin-valley physics show promise for novel applications in nano-electronics optoelectronics, flexible electronics, and quantum information processing.[10–14] Despite strong interest from industry, the 'lab-to-fab' transition faces key challenges in integrating scalable dielectrics.[2–4] While 'scotch tape exfoliation' sparked the discovery of more than a hundred 2D materials, compatible insulators remain lacking.[15–17] Hexagonal boron nitride (hBN) is commonly used but unsuitable as gate dielectrics in ultra-scaled logic or quantum devices due to its poor scalability, incompatibility with complementary metal-oxide-semiconductor (CMOS) fabrication processes, high leakage current, small dielectric permittivity and lack of zero nuclear spin isotopes.[18–22] Mechanically exfoliated hBN is inherently non-scalable with crystals that are non-uniform in size, thickness, and shape, while CVD and solution-exfoliated hBN films show poor performance when thinned to an EOT of a few nms.

One approach to scaling CMOS technology is high-$k$ (i.e., $k > k_{SiO_2}$) dielectrics like $HfO_2$ and $Al_2O_3$. For example, Intel's 14-nm FinFET structures use $HfO_2$ as a gate dielectric with equivalent oxide thickness (EOTs) of 0.9 nm.[23] However, top-down deposition techniques such as atomic layer deposition (ALD) do not integrate well with 2D TMDCs and can damage/change the underlying 2D materials.[15,17] The chemically inert surfaces of 2D materials due to the lack of dangling bonds inhibit uniform growth, leading to lower interface and dielectric qualities. Non-conformal dielectric growth with pinholes increases both interface roughness and defect densities, resulting in devices with low carrier mobilities, poor electrostatic gate control, high gate leakage currents, and permeability to potentially damaging environments. These challenges limit progress in 2D TMDC FETs; we urgently need alternative, scalable high-$k$ dielectrics compatible with 2D materials for applications in nanoelectronics, optoelectronics, valleytronics, quantum information processing, neuromorphic computing, sensing, and interconnect technologies.[6,7]

Methods for depositing high-*k* dielectrics on 2D materials using seeding layers or surface functionalization in ALD have been reported, with varying success in dielectric and interface quality and preserving 2D material properties.[15,17,24–29] Non-toxic liquid metals (LM) at room temperature are emerging candidates for synthesizing insulating ultrathin metal oxides.[30,31] These ultrathin oxides can be excellent insulating layers for 2D semiconductors while enabling additional material deposition to tailor device properties, without altering the properties of the underlying 2D materials. Such ultrathin oxidized films on LM surfaces can be easily printed onto substrates due to weaker interactions with underlying LMs, unlike in conventional mechanical exfoliation of solid metal oxides with strong interlayer bonding. By exfoliating onto polydimethylsiloxane (PDMS), ultrathin oxides can be deterministically printed onto varied materials and substrates. However, manual transfer techniques produce unwanted tears, wrinkles, excess polymer residue, and cause potential damage to either the films or to the underlying 2D materials.[31–34] To achieve high-quality integration with randomly positioned exfoliated 2D materials, we require a direct and deterministic printing technique.

Here we report the direct printing of amorphous ultrathin $Ga_2O_3$ films, a promising high-*k*, wide band gap material with excellent thermal and chemical stability.[35] After creating high-quality $Ga_2O_3$/2D material heterostructures with atomically smooth interfaces, we found valley polarization to be preserved in 2D $WS_2$ after $Ga_2O_3$ encapsulation, critical for valleytronics and optoelectronics. To investigate its use in gate dielectric stacks, we fabricated top-gated CVD monolayer $WS_2$ FETs with ~1 nm equivalent oxide thickness (EOT) using LM-printed ultrathin $Ga_2O_3$ and subsequent ALD of $HfO_2$. We recorded current on/off ratios of up to $10^5$ and excellent gate control with subthreshold swings down to 84.9 mV/dec in our FETs, while gate leakage currents measured at $<2.4\times10^{-4}$ $A/cm^2$ were up to 8 orders below those for hBN dielectrics of similar EOTs, and more than an order of magnitude lower than the requirements set out by the International Technology Roadmap for Semiconductors (ITRS) for ultra-scaled logic devices.[36] Our approach is scalable, compatible with conventional semiconductor fabrication techniques and meets thermal budget considerations for back-end-of-line (BEOL) processes in integrated circuit fabrication (~450 °C).[2,3]

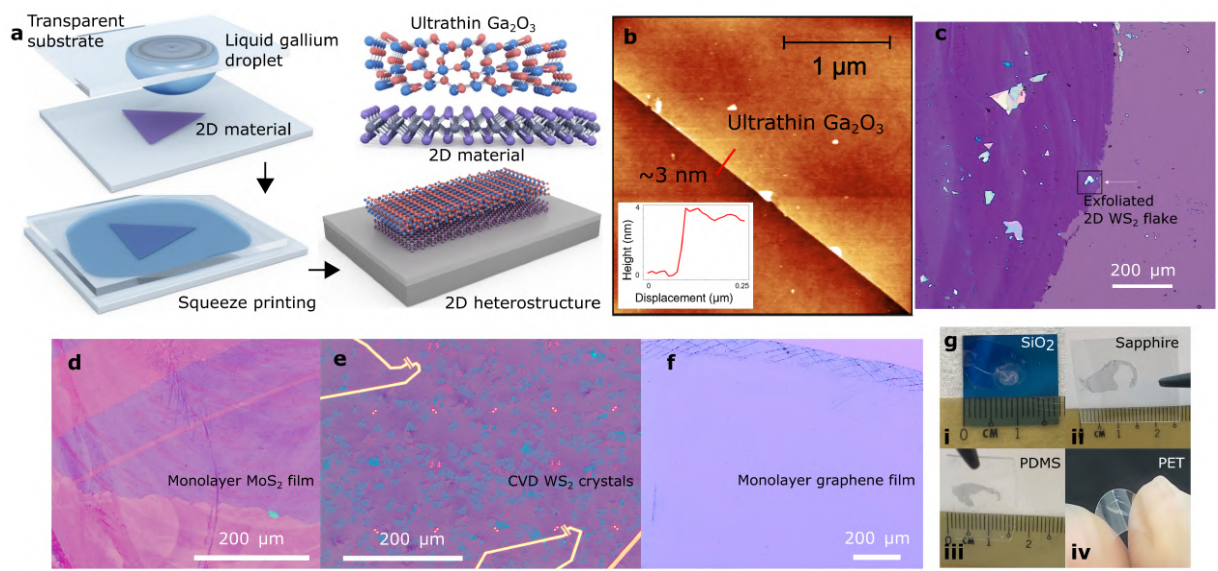

*Figure 1 **Ultrathin Ga$_2$O$_3$ printing.** (a) Schematic of our motorized ultrathin Ga$_2$O$_3$ squeeze printing process. A liquid gallium metal droplet of size ~1-3 mm is placed on a transparent sapphire substrate attached to a micromanipulator arm for precise positioning of the droplet over the target substrate e.g., close to a 2D transition metal dichalcogenide (TMDC) crystal. Substrates are brought controllably together by the motorized stage, squeezing the liquid metal droplet and conformally attaching the surface gallium oxide to the substrate surfaces. (b) Atomic force micrograph reveals a uniform and smooth Ga$_2$O$_3$ film ~3 nm thick, determined by the thickness profile (inset) at the indicated red line. Heterostructures are created by printing on 2D materials including (c) randomly positioned micrometre-sized mechanically exfoliated 2D WS$_2$, (d) large-area continuous MoS$_2$ polycrystalline monolayer films, (e) chemical vapor deposition grown isolated WS$_2$ monolayer single crystals, and (f) large-area continuous graphene films. Large-scale synthesis is possible on various substrates, as shown in the (g) camera image of cm-sized ultrathin gallium oxide on (i) SiO$_2$, (ii) sapphire, (iii) PDMS, and (iv) PET.*

## Ultrathin Ga$_2$O$_3$ printing

Drawing inspiration from the vertical stacking of 2D materials,[37,38] we developed the motorized, direct, and deterministic squeeze printing of ultrathin Ga$_2$O$_3$ films onto different substrates; and over various 2D materials (Fig. 1a). Our ability to deterministically position the liquid gallium droplet over a target area before squeeze printing ultrathin Ga$_2$O$_3$ is crucial for creating heterostructures with mechanically exfoliated or chemical vapor deposition grown (CVD) 2D materials where micron-sized flakes are randomly distributed. A motorized setup enables precise control over the printing speeds of substrates, relative tilt angles, and LM temperatures. Control over vertical and lateral shear forces during printing is important for producing high-quality films. Films produced from this motorized process are consistently more uniform and cleaner with reduced cracks, tears, wrinkles, and residues compared to manual printing or transfer. Damage to underlying 2D materials is also limited in heterostructures as shown in Supporting Information (SI) Fig. S1.

The formation of ultrathin oxides on liquid metal surfaces follows the Cabrera-Mott process ensuring a uniform film.[39] To verify the thickness and morphology of $Ga_2O_3$ films, we performed atomic force microscopy (AFM) which shows a uniform and continuous film ~3 nm thick (Fig. 1b). Next, we demonstrated direct printing of ultrathin $Ga_2O_3$ films over 2D materials, including mechanically exfoliated $WS_2$ monolayer flakes (Fig. 1c), continuous CVD polycrystalline $MoS_2$ (Fig.1d), graphene monolayer films (Fig. 1e) and CVD $WS_2$ mono- and bi-layers isolated flakes (Fig. 1f). The versatility and scalability of this approach for diverse applications, e.g., flexible optoelectronics and nano-electronics, is evident through the ability to print cm-sized films onto different substrates including silicon oxide, sapphire, PDMS and flexible PET substrates (Fig. 1g). In principle, larger films with sizes approaching wafer scale are possible; our current limitation is the size of the sample stage.

We mention an important advantage: low processing temperatures (<60 °C) due to the melting point of gallium (29.8 °C); this can be decreased further with eutectic alloys like EGaIn (15.5 °C melting point) or gallinstan (-19 °C melting point). Processing temperatures are critical in nano-electronics where thermal budgets for BEOL integrated circuit fabrication are limited to ~450 °C. Our method may potentially be applied in monolithic 3D integration where multiple layers of materials and device circuits are stacked on a single wafer.[2–4] Furthermore, our method produces atomically smooth interfaces as the liquid gallium flows and conforms to surfaces during squeeze printing.

**The $Ga_2O_3$/$WS_2$ interface**

To investigate the ultrathin $Ga_2O_3$/2D material interface, we performed AFM and cross-sectional scanning transmission electron microscopy (STEM) imaging of an exfoliated $WS_2$ encapsulated by ultrathin $Ga_2O_3$ (Fig. 2a). We first verified the thicknesses of different regions (including a smooth monolayer region ~1.2 nm thick; Fig. 2b) with AFM. Next, we examined the interfaces between $WS_2$ monolayer and few-layer regions and ultrathin $Ga_2O_3$ with cross-sectional STEM imaging (Fig. 2c, d). These images confirm that direct printing produces atomically smooth interfaces, which is critical for reducing interface roughness scattering that lowers carrier mobilities in 2D FETs.[40]. High-resolution cross-sectional STEM images (Fig. 2f) highlight the interface improvement over conventional ALD growth. We observe sub-nm separation for the LM-printed $Ga_2O_3$/$WS_2$ interface compared to the few-nm voids that line the ALD $HfO_2$/$WS_2$ interface. This solves a long-standing challenge for high-*k* dielectric integration in 2D materials where the chemical inertness of 2D material surfaces can lead to poor quality dielectric growth with atomically rough interfaces.

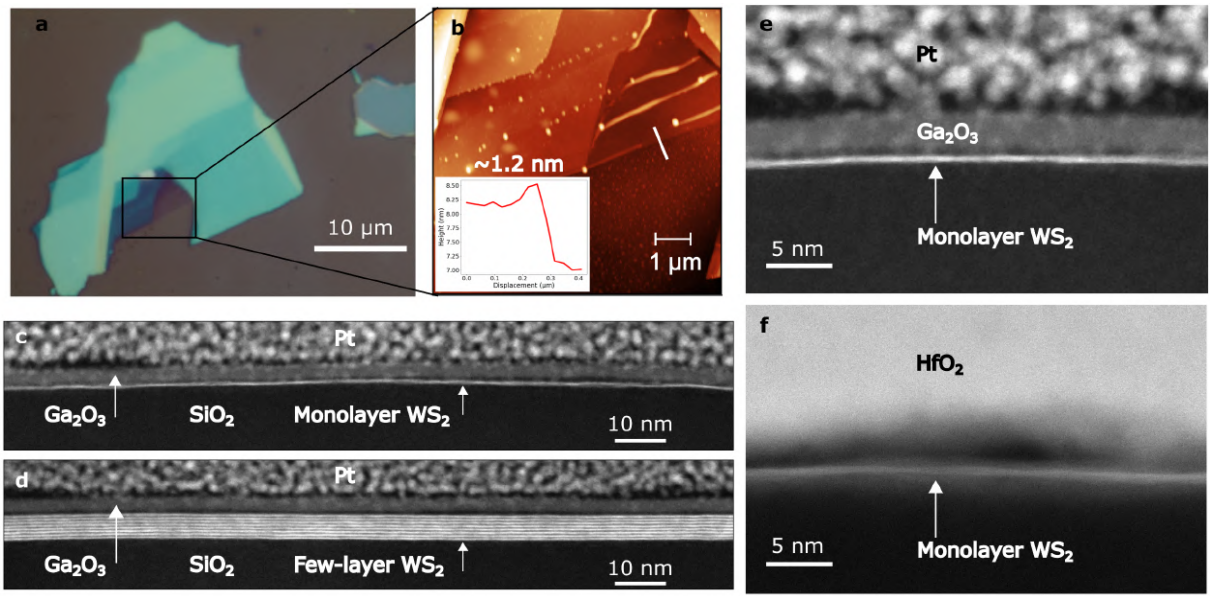

Figure 2 **Ga$_2$O$_3$/WS$_2$ interface.** (a) Optical image of exfoliated WS$_2$ crystal encapsulated by ultrathin Ga$_2$O$_3$ in Fig. 1c. (b) Atomic force micrograph reveals monolayer and few-layer regions. The thickness profile (inset) highlights a Ga$_2$O$_3$/monolayer WS$_2$ region ~1.2 nm thick. Cross-sectional scanning transmission electron microscopy images of Ga$_2$O$_3$/WS$_2$ interface for (c) monolayer and (d) few-layer region of the WS$_2$ in (a). (e) Close-up of Ga$_2$O$_3$/monolayer WS$_2$ region, showing an atomically smooth interface with sub-nm separation between Ga$_2$O$_3$ and WS$_2$. In contrast, the interface of an (f) atomic layer deposition grown HfO$_2$/monolayer WS$_2$ region shows an irregular and rough interface with inter-layer separation of a few nm.

### Valley polarization of Ga$_2$O$_3$/WS$_2$

Another interesting question is the suitability of Ga$_2$O$_3$ for optoelectronics and valleytronics.[41] Ga$_2$O$_3$ is an isotropic insulator, and being transparent across the visible spectrum, it is potentially ideal for optoelectronics. However, its usability in valleytronics in conjunction with 2D materials such as WS$_2$ remains unknown. Monolayer WS$_2$ has a unique spin-valley coupling and optical selection rules from its band structure with two non-equivalent symmetry points $K$ and $K'$. Valley polarization from carrier population imbalance between the K and K' valleys is detectable through circular dichroic photoluminescence spectroscopy (CDPL).[42] We investigated spin-valley polarization through CDPL mapping of CVD grown monolayer WS$_2$ crystals with and without Ga$_2$O$_3$ cover (Fig. 3f-h) at $T$=3.6 K. Notably, the degree of circular polarization (DoCP) in 2D WS$_2$ remained homogeneous with no significant changes between regions covered and uncovered by Ga$_2$O$_3$ (Fig. 3c, f), indicating preservation of spin-valley coupling. These measurements suggest LM-printed Ga$_2$O$_3$ to be an excellent dielectric for 2D TMDC-based applications in optoelectronics and valleytronics. Our technique for creating Ga$_2$O$_3$/WS$_2$ heterostructures allows further investigations and insights into the sensitivity of intervalley scattering in 2D WS$_2$ to its immediate environment.

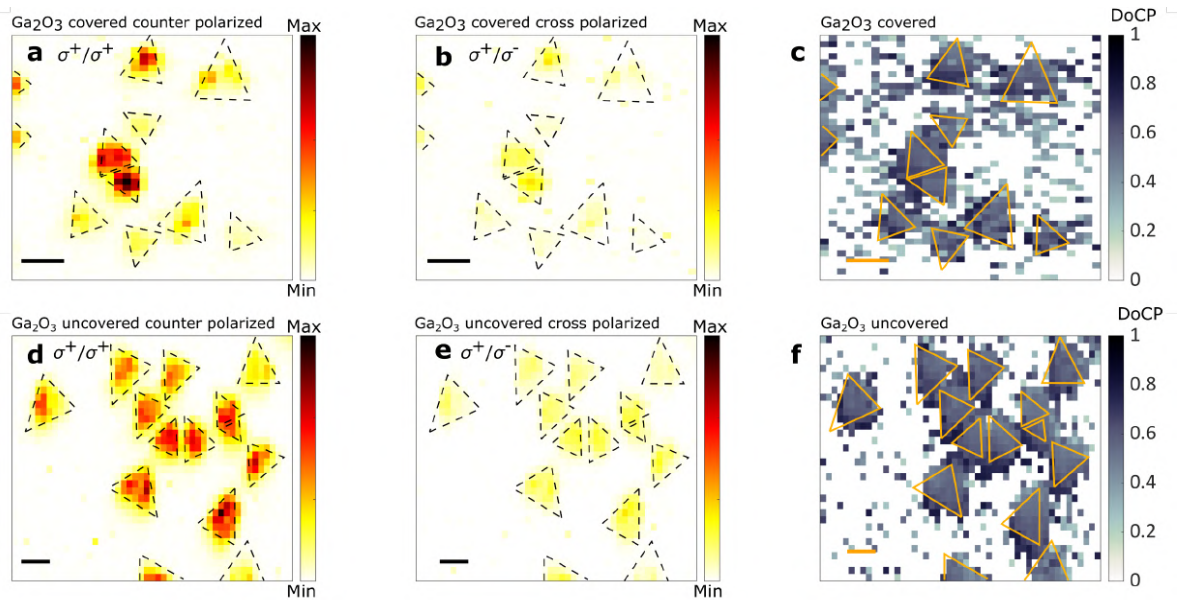

*Figure 3 **Circular polarization in $Ga_2O_3$/$WS_2$ heterostructures.** Photoluminescence mapping at T=3.6 K of regions with chemical vapor deposition grown monolayer $WS_2$ flakes (a, b) covered and (d, e) uncovered by $Ga_2O_3$. Co-polarization photoluminescence is shown in (a, d) and cross-polarization in (b, e). Boundaries of monolayer flakes are delineated with black dashed lines for clarity. The homogeneous distribution of the degree of circular polarization (DoCP) can be observed for both (with borders shown in yellow lines) (c) covered and (f) uncovered regions with DoCP ~0.4-0.5. Scale bars are 5 μm.*

### **Dual-gated $HfO_2$/$Ga_2O_3$/$WS_2$ transistor**

An urgent challenge in integrated circuits of 2D semiconductors such as $WS_2$ is the scalable, damage-free integration of compatible dielectrics for top gates without compromising intrinsic 2D semiconductor properties. Ideally, gate dielectrics should be compatible with pinhole-free, conformal deposition techniques like ALD for sub-nanometre wafer-scale thickness control. We fabricated dual-gated CVD-grown monolayer $WS_2$ FETs with LM-printed ultrathin $Ga_2O_3$/ALD $HfO_2$ as top-gate dielectrics and $SiO_2$/Si substrate as back gates (Fig. 4a, b). We hereafter focus our discussion on detailed measurements of two devices with different gate and channel widths (Fig. 4c). Transport characteristics of 3 more devices are available in the SI. The excellent protective capability of $Ga_2O_3$ against $O_2$ plasma and further material deposition; and its suitability for promoting ultrathin $HfO_2$ ALD growth are shown from AFM and electrical measurements found in the SI.

First, to understand the influence of $Ga_2O_3$ on $WS_2$ carrier mobility, we measured the two-terminal carrier mobility of a CVD monolayer $WS_2$ device covered by ultrathin $Ga_2O_3$ (Fig. 4d). At temperatures >200 K, the transport behavior is consistent with previous reports of mobility scaling $T^{-\gamma}$ from increased electron–phonon scattering dominant at elevated temperatures. Adding $Ga_2O_3$ on top of $WS_2$ mechanically quenches the out-of-plane homopolar mode, leading to a smaller exponent γ=0.78 compared to ~1.5-1.9 for uncovered 2D $WS_2$.[40]

The quality of our top-gate $Ga_2O_3$/$HfO_2$ dielectric stack (thicknesses ~3 nm for LM-printed $Ga_2O_3$ and ~3 nm for ALD $HfO_2$) can be observed from the transport characteristics. An important figure of merit for assessing efficient channel control is a low subthreshold swing (SS) approaching the thermionic limit of 60 mV/dec, which can be limited in 2D TMDC devices due to poor interface quality. We determined SS from drain currents as a function of top-gate voltages and found low values for both devices [84.9 mV/dec (device A, Fig. 4e) and 121.7 mV/dec (device B, Fig. 4f)]. Double transfer curve sweeps at different drain voltages show effective hysteresis-free behavior with current on/off ratio ~$10^5$ at various drain voltages (Fig. 4g, h). The small SS values and hysteresis-free transfer characteristics show that direct LM-printing affords excellent interface properties and a small EOT. This contrasts with manually transferred $Ga_2O_3$ films where large hysteresis loops due to interface trapped charges were measured.[33] Furthermore, room temperature drain currents exhibit saturations at $V_D$<1 V from channel pinch-off which are consistent with small EOT devices (Fig. 4i, j).[24]

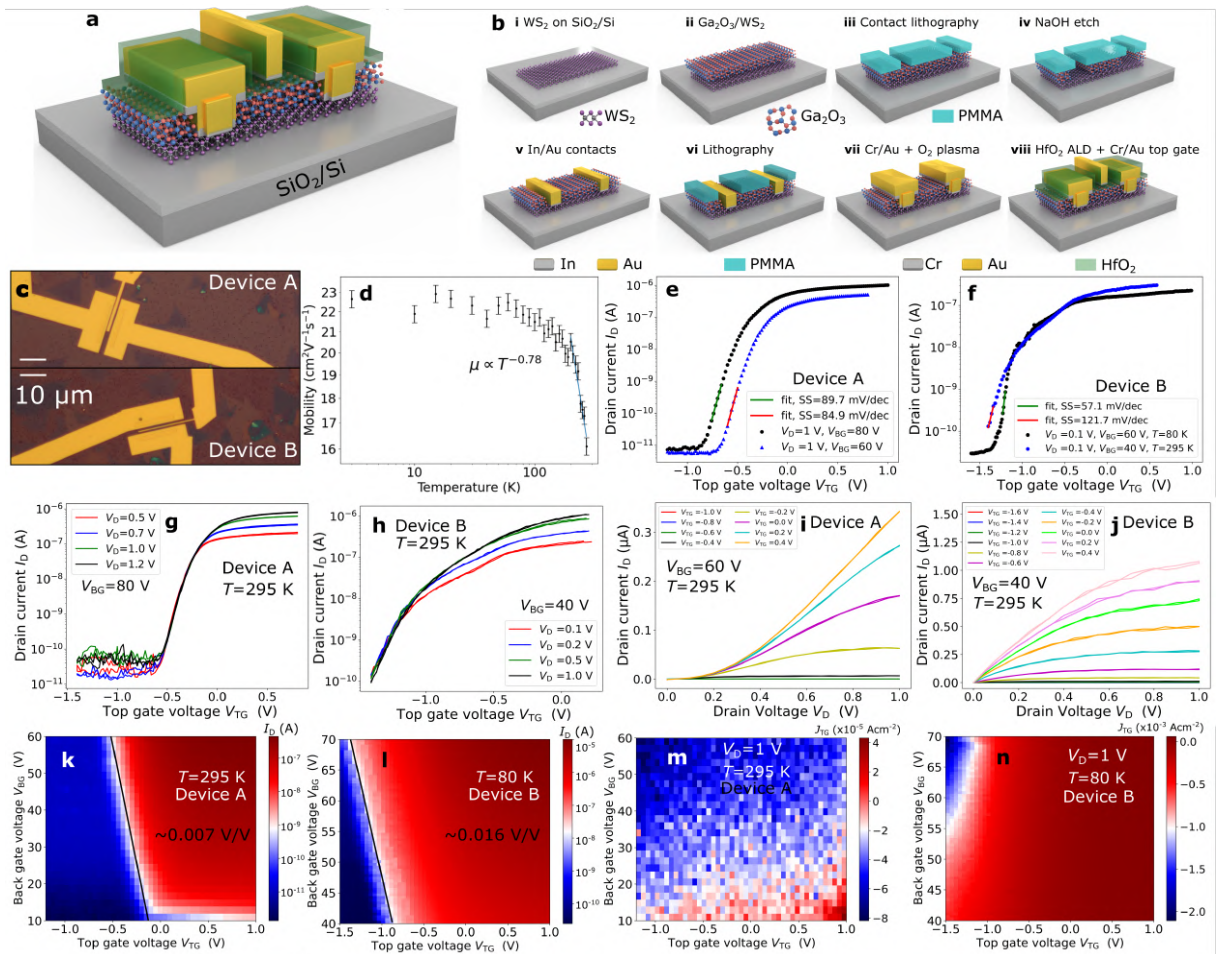

*Figure 4 **Electrical measurements of dual-gated 2D WS₂ transistor.** (a) Schematic of our 2D WS₂ device with ~6 nm Ga₂O₃/HfO₂ top-gate dielectric stack. The fabrication process is illustrated in (b). (c) Optical images of two typical 2D WS₂ devices A (gate length=1.2 μm, channel width=7.5 μm) and B (gate length=0.5 μm, channel width=8 μm). (d) Two-terminal back-gated carrier mobility of a CVD monolayer WS₂/Ga₂O₃ transistor. Top-gated transfer curves of (e) device A and (f) device B. Red and green lines represent fits to determine subthreshold swings. Double sweep top-gated transfer curves at different drain voltages are shown for (g) device A and (h) device B, showing negligible hysteresis. Output curves at different top-gate voltages for (i) device A and (j) device B. Drain current as a function of back-gate and top-gate voltages at drain voltage $V_D$=1 V for (k) device A and (l) device B. Black lines represent fits to determine lever arms between top and back-gates. Top-gate leakage current densities as a function of back-gate and top-gate voltages at drain voltage $V_D$=1 V for (m) device A and (n) device B.*

We estimated EOTs with dual-gate measurements of the drain currents as a function of top- and back-gate voltages, which show a lever arm of ~0.007 V/V (device A, Fig. 4k) and ~0.016 V/V (device B, Fig. 4l). We find small EOT estimates of 0.60-1.2 nm (device A) and 0.57-1.1 nm (device B) (calculation details are found in SI) consistent with the high-*k* nature of our top-gate dielectric stack. Improving device fabrication should overcome laboratory limitations in lithographic alignment and Ga₂O₃ wet etch for increased top-gate capacitance, potentially further reducing SS to approach the thermionic limit of 60 mV/dec.

With the EOTs, we calculated interface trap densities $D_{IT}$ from SS=$\frac{2.3k_BT}{q}(1+\frac{qD_{IT}}{C_{OX}})$, where $q$ is the electron charge, $k_B$ the Boltzmann constant, and $C_{OX}$ the gate oxide capacitance. $D_{IT}$ = 4.8x10¹² cm⁻²eV⁻¹ (device A) and 4.9x10¹² cm⁻²eV⁻¹ (device B). Contributions to $D_{IT}$ can be from defects at insulator surfaces, e.g., oxide dangling bonds, or channel defects. The latter is dominant in devices with CVD flakes, which are more defective than exfoliated flakes, and likely the limiting factor in our devices.[15] For example, ALD HfO₂-capped growth yielded SS = 60 mV/dec when using exfoliated MoS₂, but only 160 mV/dec when using CVD-grown MoS₂ due to increased channel defect density.[24] Our $D_{IT}$ values are consistent with similar reports for CVD-based devices. Furthermore, these values are lower than control devices we made of WS₂ mono- and bi-layers without the top encapsulating layer (SI Fig. S5 and S6), suggesting low contributions to trap densities from Ga₂O₃/WS₂ interfaces.

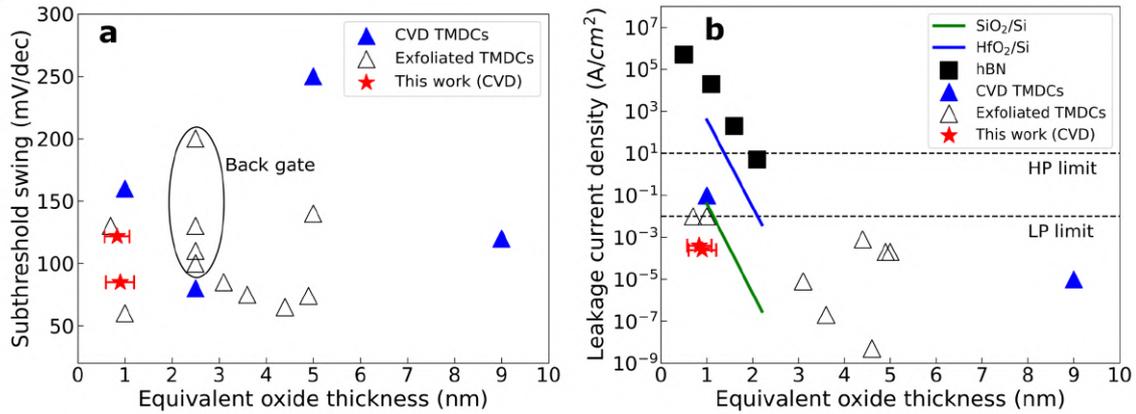

*Figure 5 **Benchmarking of subthreshold swings and gate leakage.** (a) Subthreshold swings vs equivalent oxide thicknesses compared with other 2D semiconductor field effect transistors (benchmarked against chemical vapor deposition (CVD) grown and exfoliated TMDC devices). Devices with back gates are highlighted. (b) Leakage current densities vs equivalent oxide thicknesses in comparison with other dielectric approaches for 2D semiconductors, including the commonly used hBN, and mature Si technologies. Horizontal dashed lines represent limits for high-performance (HP) and lower-power (LP) logic devices as defined by the International Technology Roadmap for Semiconductors.[36]*

Another important requirement for efficient low-power circuits is a gate leakage current <0.01 A/cm$^2$, especially in ultra-scaled FETs where gate leakage currents increase with decreasing oxide thickness.[43] A major limitation of hBN arising from its low dielectric permittivity is an excessive gate leakage current >10 Acm$^{-2}$ for crystalline monolayers when scaled to ~1 nm EOT.[18] For our devices, we find excellent gate leakage currents $J_{TG}$ with $J_{TG}$<2.5x10$^{-4}$ A/cm$^2$ (device A) and <4.0x10$^{-4}$ A/cm$^2$ (device B) at $V_{TG}$=-1.6 V and a large drain voltage $V_D$ of 1 V (SI Fig. S9). Comprehensive measurements of top-gate leakage currents as a function of back- and top-gate voltages are shown in Fig. 4m, n).

**<u>Benchmarking</u>**

Finally, we assessed industrial potential by benchmarking our devices against other reported studies on 2D semiconductors and mature Si technologies. Because of the paucity of reports on CVD monolayer WS$_2$, we also include exfoliated and CVD mono- and multi-layer MoS$_2$ in our benchmarking (SI Table 1). Our devices rank among the lowest SS and EOT values, comparable even to exfoliated multi-layered MoS$_2$ flakes with far lower channel defect densities, highlighting the excellent Ga$_2$O$_3$/WS$_2$ interface characteristics (Fig. 5b). Even at 1 nm EOT, our devices display gate leakage currents lower than mature SiO$_2$/Si and HfO$_2$/Si technologies, and more than 8 orders of magnitude lower than those using hBN (Fig. 5c). Our gate leakage currents easily meet requirements for CMOS low power circuits (<0.01 Acm$^{-2}$) as projected by the ITRS report.[36]

**Conclusion**

In conclusion, we demonstrated a promising approach for integrating high-*k* dielectrics on 2D materials with atomically smooth interfaces fabricated by direct LM printing of ultrathin $Ga_2O_3$. Our method preserved valley polarization in 2D $WS_2$, highlighting its potential for use in optoelectronics and valleytronics. We successfully produced a top-gated CVD monolayer $WS_2$ FET with a ~1 nm EOT gate stack, excellent subthreshold swings, atomically smooth interfaces, and low gate leakage currents using LM-printed $Ga_2O_3$/ALD $HfO_2$ gate stack. Furthermore, ultrathin oxides beyond $Ga_2O_3$ have been successfully demonstrated in other eutectic alloys.[44] The theoretical prediction of other viable oxides can potentially generate a library of ultrathin, scalable materials for 2D heterostructures with novel functionalities. Our technique is scalable, low cost, and compatible with BEOL thermal budgets and conventional semiconductor fabrication techniques. There are several opportunities for further exploration and optimization e.g., improved printing recipes or post-processing treatments such as annealing to tune the ultrathin oxide properties. Developing alternative compatible dielectrics solves a key challenge en-route to the industrial use of 2D semiconductors.

**Methods**

**CVD growth and transfer of $WS_2$:** In a typical growth process, $WO_2$ (99.9%, Sigma-Aldrich) and sulfur powders (99.999%, Sigma-Aldrich) were used as the metal and the chalcogen sources, respectively. The sapphire (Princeton Scientific Corporation) was annealed at 1100 °C for 2 hours before use as a growth substrate. A quartz crucible containing 15 mg of $WO_2$ powder was placed at the center of the furnace. The annealed sapphire was mounted on top of the crucible with its polished surface facing downwards. A crucible filled with 1600 mg of sulfur powder was placed upstream and heated to a temperature of 250 °C. The atmospheric pressure growth was performed in a tube furnace at 900 °C under 30 sccm of Ar gas. After 15 min growth, the furnace was naturally cooled under an Ar flow of 200 sccm. Grown flakes were wet transferred onto a $SiO_2$/Si substrate that also operates as a back gate. Poly(methyl methacrylate) (PMMA) was spin-coated on the surface of the sapphire substrate and the $WS_2$/PMMA stack was detached with PDMS and DI water and transferred onto the $SiO_2$/Si substrate. The substrate was heated to 120 °C to remove the PDMS stamp, then dried overnight before PMMA removal with acetone and Microposit Remover 1165 solvent.

**Liquid metal oxide printing:** First, we positioned a liquid gallium droplet of size ~1-3 mm on the top transparent substrate (typically sapphire) which was attached to a micromanipulator with multi-axis control under ambient conditions. With an optical microscope and high-resolution micromanipulator, we precisely positioned the liquid metal droplet over the target area where the ultrathin oxide film was to be printed, e.g., near a 2D transition metal dichalcogenide crystal. The bottom target substrate on a motorized stage was heated to 60 °C and slowly brought into contact with the liquid gallium droplet. The relative position between the substrates was slowly reduced (~1 µm/s) for the liquid metal to thermally equilibrate. The liquid metal droplet was 'squeezed' and expanded across the substrate surfaces, with the ultrathin oxide skin attached to the substrates. Finally, the substrates were withdrawn at a rate of ~10 µm/s. Excess liquid gallium residue was mechanically cleaned by immersing the substrate in boiling ethanol and wiping it off with a clean, soft, lint-free cotton bud. Finally, a PDMS film was wiped over the ethanol submerged film to remove any remaining liquid metal residue.

**Device fabrication:** Contact electrodes were patterned using standard e-beam lithography and PMMA resist. After development, we selectively wet etched the $Ga_2O_3$ films with a 0.1 M solution of NaOH for 20 s. Following this, 5/40 nm of In/Au was deposited at 0.1 Å/s followed by liftoff in Microposit remover 1165 solution. Next, we patterned an overlap contact structure designed to protect the devices against $O_2$ plasma due to side wall etching during the NaOH etch step. 10/50 nm of Cr/Au was deposited. We then performed $O_2$ plasma treatment at 50 W, 20 sccm $O_2$, and 200 mTorr for 5 s to clean the $Ga_2O_3$ surface from resist residues and promote its hydrophilicity for subsequent ALD $HfO_2$ growth. The $HfO_2$ was deposited with an Anric AT410 system attached to a glove box using TDMA-HF and $H_2O$ as precursors at 200 °C for 30 cycles. Top-gate electrodes were fabricated with standard electron beam lithography. 5/40 nm of Cr/Au was deposited at 0.1 Å/s followed by liftoff in Microposit remover 1165 solution.

**Optical measurements:** Room temperature PL and Raman spectra were obtained with a 532 nm laser with objective x100 (NA = 0.85) and 2400 l/mm grating. For low-temperature CDPL mapping measurements, the sample was kept in a cryostat on top of a motorized stage and a 570 nm pulsed laser (~80 ps) was used for the excitation with a x50 objective (NA = 0.5) used for the detection. This 570 nm pulse was generated from NKT supercontinuum source by placing a bandpass filter with a bandwidth of 6 nm. The excitation power was kept at ~15 µW to avoid any unintentional degradation due to laser-induced heating. The degree of circular polarization (DoCP) was calculated as: $\text{DoCP} = \frac{I_{co} - I_{cross}}{I_{co} + I_{cross}}$ where $I_{co}$ is the intensity of the co-polarized light and $I_{cross}$ the intensity of the cross polarized light.

**Electrical measurements:** Electrical measurements were performed in a cryogenic probe station using Keithley 2450 SMUs.

**STEM imaging:** Cross-sectional samples were prepared with focused ion beam milling using a ThermoFisher Helios 450S dual-beam NanoLab. Final milling was done at with gentle beam conditions at 2 keV. Aberration-corrected STEM was performed with a JEOL ARM200 CFEG, equipped with a CEOS DCOR aberration-corrector, and operated at 80 kV. The presented images were acquired in the high-angle annular dark field detection mode, giving bright contrast in areas of high atomic number (such as $WS_2$) and dark contrast in areas of low atomic number (such as $SiO_2$).

**X-ray photoemission spectroscopy:** XPS data were acquired using an Al K$\alpha$ source (photon energy $h\nu$=1486.7 eV) and a beam spot size of about 200 µm with an energy resolution of $\approx$ 0.3 eV. The photoelectrons were collected at normal emission angle and the light was incident at 60° to the surface normal. Binding energies were calibrated against the Au 4f core level energy of a gold reference sample. All XPS data were acquired at 300 K.

**Density functional theory calculations:** All DFT-based calculations were carried out using the Vienna ab Initio Simulation Package (VASP). The Projector augmented wave pseudopotentials for the core and the Perdew-Burke-Ernzerhof (PBE) format of the generalized gradient approximation (GGA) for the exchange-correlation functional were adopted. We set the kinetic energy cutoff to be 450 eV, and optimized the structures until the forces were below 0.01 eV/Å.

**Acknowledgments**

This research was supported by the Agency for Science, Technology, and Research (A*STAR) under its MTC YIRG grant No. M21K3c0124. We acknowledge the funding support from Agency for Science, Technology and Research (#21709). F.B., I.V. and K.E.J.G. acknowledges support from a Singapore National Research Foundation Grant (CRP21-2018-0094). C.P.Y.W. acknowledges funding support from the A*STAR AME YIRG Grant No. A2084c0179. Z.L. and Y.S.A. are supported by Singapore Ministry of Education (MOE) Academic Research Fund Tier 2 (MOE-T2EP50221-0019) and A*STAR AME IRG (No. A2083c0057). M.B. acknowledges support from the Ministry of Education (MOE) Singapore, under AcRF Tier 1 startup Grant No. R-284-000-179-133 / A-0009238-01-00. H.D. is supported by A*STAR under its CDF-SP award C222812022.

**Contributions**

C.S.L., Y.Y.Z., D.V., R.L. and fabricated the devices. Z.J. and C.P.Y.W assisted with liquid metal oxide printing. F.W. and S.W.T. grew the materials. C.S.L., Y.Y.Z, and D.V. performed and analyzed the electrical measurements. S.D., D.H., I.V., and O.Z.E. performed and analyzed the CDPL experiments. M.B. and S.L.T. performed cross-sectional STEM imaging. J.W.C. assisted with ALD. F.B. performed and analyzed the XPS measurements. Z.L.J. and Y.S.A. performed the DFT calculations. K.E.J.G. and C.S.L. supervised the experimental work. C.S.L. conceived and directed the research. C.S.L. wrote the manuscript with input from all authors.